# *Generalization Strategies for the Verification of Infinite State Systems*

FABIO FIORAVANTI

*Dipartimento di Scienze, Università 'G. D'Annunzio',*
*Viale Pindaro 42, I-65127 Pescara, Italy*
(*e-mail:* `fioravanti@sci.unich.it`)

ALBERTO PETTOROSSI

*Dipartimento di Informatica, Sistemi e Produzione,*
*Università di Roma Tor Vergata, Via del Politecnico 1, I-00133 Roma, Italy*
(*e-mail:* `pettorossi@disp.uniroma2.it`)

MAURIZIO PROIETTI

*CNR-IASI, Viale Manzoni 30, I-00185 Roma, Italy*
(*e-mail:* `maurizio.proietti@iasi.cnr.it`)

VALERIO SENNI

*LORIA-INRIA, 615, rue du Jardin Botanique*
*BP 101, 54602 Villers-les-Nancy Cedex, France,*
*& Dipartimento di Informatica, Sistemi e Produzione,*
*Università di Roma Tor Vergata, Via del Politecnico 1, I-00133 Roma, Italy*
(*e-mail:* `valerio.senni@loria.fr`, `senni@disp.uniroma2.it`)

## Abstract

We present a method for the automated verification of temporal properties of infinite state systems. Our verification method is based on the specialization of constraint logic programs (CLP) and works in two phases: (1) in the first phase, a CLP specification of an infinite state system is specialized with respect to the initial state of the system and the temporal property to be verified, and (2) in the second phase, the specialized program is evaluated by using a bottom-up strategy. The effectiveness of the method strongly depends on the generalization strategy which is applied during the program specialization phase. We consider several generalization strategies obtained by combining techniques already known in the field of program analysis and program transformation, and we also introduce some new strategies. Then, through many verification experiments, we evaluate the effectiveness of the generalization strategies we have considered. Finally, we compare the implementation of our specialization-based verification method to other constraint-based model checking tools. The experimental results show that our method is competitive with the methods used by those other tools. To appear in Theory and Practice of Logic Programming (TPLP).

## 1 Introduction

We consider the problem of verifying properties of reactive systems, that is, systems which continuously react to inputs by changing their internal state and producing outputs. One of the most challenging problems in this area is the extension of the model checking technique (Clarke et al. 1999) from finite state systems to infinite state systems. In infinite state model checking the evolution over time of a system is modelled as a binary transition relation on an infinite set of states and the properties of that evolution are specified by means of propositional temporal formulas. In particular, in this paper we consider the *Computation Tree Logic* (CTL), which is a branching time propositional temporal logic by which one can specify, among others, the so-called *safety* and *liveness* properties (Clarke et al. 1999).

Unfortunately, the verification of CTL formulas for infinite state systems is, in general, an undecidable problem. Thus, in order to cope with this limitation, various *decidable subclasses* of systems and formulas have been identified (see, for instance, (Esparza 1997)). Other approaches to overcome the undecidability limitation are based on the enhancement of finite state model checking by using either *deductive techniques* (Pnueli and Shahar 1996; Sipma et al. 1999) or *abstractions*, by which one can compute conservative approximations of sets of states (Abdulla et al. 2009; Bultan et al. 1999; Clarke et al. 1994; Dams et al. 1997; Geeraerts et al. 2006; Godefroid et al. 2001).

Constraint logic programming (CLP) provides an excellent framework for specifying and verifying properties of reactive systems (Fribourg 2000). Indeed, the fixpoint semantics of logic programming languages allows us to easily represent the fixpoint semantics of various temporal logics (Delzanno and Podelski 2001; Nilsson and Lübcke 2000; Ramakrishna et al. 1997) and constraints over the integers or the reals can be used to provide finite representations of infinite sets of states (Delzanno and Podelski 2001; Fribourg and Olsén 1997).

However, for programs that specify infinite state systems, the proof procedures normally used in CLP, such as the extensions of SLDNF resolution and tabled resolution (Cui and Warren 2000), very often diverge when trying to check some given temporal properties. This is due to the limited ability of these proof procedures to cope with infinitely failed derivations. For this reason, instead of using direct program evaluation, many logic programming-based verification systems make use of reasoning techniques such as: (i) *abstract interpretation* (Banda and Gallagher 2010; Delzanno and Podelski 2001), and (ii) *program transformation* (Fioravanti et al. 2001; Leuschel and Lehmann 2000; Leuschel and Massart 2000; Peralta and Gallagher 2003; Roychoudhury et al. 2000). In the techniques based on abstract interpretation one can construct approximations of the least and greatest fixpoints of (the immediate consequence operator associated with) a CLP program and then check the properties of interest on these approximations, while in the techniques based on program transformation one can pre-process the specification of a given system and a given property so that the verification itself becomes easier to perform.

This paper presents a verification method based on *program specialization*, a transformation technique that improves a program by exploiting the knowledge about the specific context where the program is used (Jones et al. 1993; Leuschel and Bruynooghe 2002). Our verification method is an extension of the one first proposed in (Fioravanti et al. 2001) and is applicable to the specification of a CTL property of an infinite state system encoded as a CLP program with locally stratified negation, where the constraints are linear inequations over the rationals. Our verification method works in two phases. In Phase (1) we specialize the CLP program with respect to the initial state of the system and the temporal property to be verified, and in Phase (2) we construct the perfect model of the specialized program derived at the end of Phase (1), by applying a bottom-up evaluation procedure. As we will demonstrate through many examples below, this bottom-up procedure terminates in most cases without the need for abstractions.

The effectiveness of the verification method we propose, strongly depends on the design of the *generalization strategy* which has to be applied during the program specialization phase. Designing a good generalization strategy is not a trivial task: it must guarantee the termination of the specialization phase, and it should also provide a high precision and good performance. These requirements are often conflicting because, on the one hand, the use of a too coarse generalization strategy may determine the non-termination of Phase (2) and, thus, prevent the verification of many interesting properties and, on the other hand, a too specific generalization strategy may lead to verification times which are too long.

In this paper we introduce some new generalization strategies and we also propose various generalization strategies which are obtained by combining old techniques, already considered in the field of program analysis and program transformation (such as the *well-quasi orders* (Leuschel 2002; Leuschel et al. 1998; Sørensen and Glück 1995) and the *convex hull* and *widening operators* (Cousot and Halbwachs 1978; Peralta and Gallagher 2003)).

Our verification method has been implemented on the MAP transformation system (MAP 2011). We have evaluated the effectiveness of this method by presenting the results of the experiments we have performed on several infinite state systems and temporal properties. We have also compared the implementation of our verification method with the following constraint-based model checking tools: (i) ALV (Yavuz-Kahveci and Bultan 2009), (ii) DMC (Delzanno and Podelski 2001), and (iii) HyTech (Henzinger et al. 1997). The experiments we have performed show that our method is effective and competitive with respect to the methods implemented in those verification tools.

The paper is structured as follows. In Section 2 we recall how CTL properties of infinite state systems can be encoded by using locally stratified CLP programs. In Section 3 we present our two-phase verification method. In Section 4 we describe various strategies that can be applied during Phase (1), that is, the specialization phase, and in particular, the generalization strategies used for ensuring termination of that phase. In Section 5 we report on some experiments we have performed by using a prototype implemented on the MAP transformation system.

## 2 Specifying Reactive Systems and CTL Properties by CLP Programs

A reactive system is modelled as a *Kripke structure*, denoted by a 4-tuple $\langle S, I, R, L \rangle$, where $S$ is a (possibly infinite) set of *states*, $I \subseteq S$ is the set of *initial states*, $R$ is a total binary *transition relation*, and $L$ is a *labeling function* that associates with each state the set of *elementary properties* that hold in that state. A *computation path* in $\mathcal{K}$ is an *infinite* sequence of states $s_0\, s_1 \ldots$ such that, for every $i \geq 0$, $s_i\, R\, s_{i+1}$ holds. The state $s_{i+1}$ is called a *successor* of $s_i$. The properties to be verified will be specified as formulas of the *Computation Tree Logic* (CTL), whose syntax is:

$$\varphi ::= e \mid \mathit{not}(\varphi) \mid \mathit{and}(\varphi_1, \varphi_2) \mid \mathit{ex}(\varphi) \mid \mathit{eu}(\varphi_1, \varphi_2) \mid \mathit{af}(\varphi)$$

where $e$ belongs to the set *Elem* of the elementary properties. Note that, in order to be consistent with the syntax of constraint logic programs, we slightly depart from the syntax of CTL given in (Clarke et al. 1999).

The operators $ex, eu$, and $af$ have the following semantics. The formula $ex(\varphi)$ holds in a state $s$ if there exists a successor $s'$ of $s$ such that $\varphi$ holds in $s'$. The formula $eu(\varphi_1, \varphi_2)$ holds in a state $s$ if there exists a computation path $\pi$ starting from $s$ such that $\varphi_1$ holds in all states of a finite prefix of $\pi$ and $\varphi_2$ holds in the first state of the rest of the path. The formula $af(\varphi)$ holds in a state $s$ if on every computation path $\pi$ starting from $s$ there exists a state $s'$ where $\varphi$ holds. Formally, the semantics of CTL is given by the satisfaction relation $\mathcal{K}, s \models \varphi$, which tells us when a formula $\varphi$ holds in a state $s$ of the Kripke structure $\mathcal{K}$.

All CTL operators can be defined in terms of $ex, eu$, and $af$. For instance: (i) the formula $ef(\varphi)$ (which holds in a state $s$ if there exists a computation path $\pi$ starting from $s$ and a state on $\pi$ where $\varphi$ holds) is defined as $eu(true, \varphi)$, and (ii) the formula $eg(\varphi)$ (which holds in a state $s$ if there exists a computation path $\pi$ starting from $s$ such that, for every state on $\pi$, $\varphi$ holds) is defined as $not(af(true, \varphi))$.

In order to encode a Kripke structure and the satisfaction relation as a CLP program we will consider a set $\mathcal{C}$ of constraints and an interpretation $\mathcal{D}$ for the constraints in $\mathcal{C}$. We assume that: (i) $\mathcal{C}$ contains a set of *atomic constraints*, among which are *true*, *false*, and the equalities between terms, denoted by $t_1 = t_2$, (ii) $\mathcal{C}$ is closed under *conjunction* (denoted by comma), and (iii) $\mathcal{C}$ is closed under *projection*. The projection of a constraint $c$ onto a tuple $X$ of variables, denoted $\mathit{project}(c, X)$, is a constraint such that $\mathcal{D} \models \forall X \, (\mathit{project}(c, X) \leftrightarrow \exists Y c)$, where $Y$ is the tuple of variables occurring in $c$ and not in $X$. We define a partial order $\sqsubseteq$ on $\mathcal{C}$ as follows: for any two constraints $c_1$ and $c_2$ in $\mathcal{C}$, $c_1 \sqsubseteq c_2$ iff $\mathcal{D} \models \forall \, (c_1 \rightarrow c_2)$.

The semantics of a CLP program is defined as a $\mathcal{D}$-*model* (Jaffar and Maher 1994), that is, a (possibly infinite) set of ground atoms whose truth implies the truth of all clauses of the program. Similarly to the case of logic programs, every *locally stratified* CLP program $P$ has a unique *perfect $\mathcal{D}$-model* (also called *perfect model*, for short) which is denoted by $M(P)$ (see, for instance, (Apt and Bol 1994)).

Now, a Kripke structure $\langle S, I, R, L \rangle$ can be encoded by a CLP program as follows.
(1) A state in $S$ is encoded by an $n$-tuple $\langle t_1, \ldots, t_n \rangle$ of terms representing the values of the variables of the reactive system. In what follows the variables $X$ and $Y$ are assumed to range over $S$.
(2) An initial state $X$ in $I$ is encoded by a clause of the form $\mathit{initial}(X) \leftarrow c(X)$, where $c(X)$ is a constraint.
(3) The transition relation $R$ is encoded by a set of clauses of the form $t(X, Y) \leftarrow c(X, Y)$, where $c(X, Y)$ is a constraint. We also introduce a predicate $ts$ such that, for every state $X$, $Ys$ is a list of all the successor states of $X$ iff $ts(X, Ys)$ holds, that is, for every state $X$, the state $Y$ belongs to the list $Ys$ iff $t(X, Y)$ holds. In (Fioravanti et al. 2007) the reader will find: (i) an algorithm for deriving the clauses defining $ts$ from the clauses defining $t$, and (ii) some conditions that guarantee that $Ys$ is a finite list.

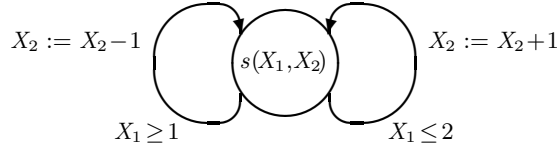

Figure 1. A reactive system. In any initial state we have that $X_1 \leq 0$ and $X_2 = 0$. The transitions do not change the value of $X_1$.

(4) Each elementary property $e$ associated with a state $X$ is encoded by a clause of the form $elem(X, e) \leftarrow c(X)$, where $c(X)$ is a constraint.

The satisfaction relation $\models$ can be encoded by a predicate $sat$ defined by the following clauses (Fioravanti et al. 2001) (see also (Leuschel and Massart 2000; Nilsson and Lübcke 2000) for similar encodings):

1. $sat(X, F) \leftarrow elem(X, F)$
2. $sat(X, not(F)) \leftarrow \neg sat(X, F)$
3. $sat(X, and(F_1, F_2)) \leftarrow sat(X, F_1), sat(X, F_2)$
4. $sat(X, ex(F)) \leftarrow t(X, Y), sat(Y, F)$
5. $sat(X, eu(F_1, F_2)) \leftarrow sat(X, F_2)$
6. $sat(X, eu(F_1, F_2)) \leftarrow sat(X, F_1), t(X, Y), sat(Y, eu(F_1, F_2))$
7. $sat(X, af(F)) \leftarrow sat(X, F)$
8. $sat(X, af(F)) \leftarrow ts(X, Ys), sat\_all(Ys, af(F))$
9. $sat\_all([\,], F) \leftarrow$
10. $sat\_all([X|Xs], F) \leftarrow sat(X, F), sat\_all(Xs, F)$

Suppose that we want to verify that a CTL formula $\varphi$ holds for all initial states. In order to do so we define a new predicate $prop$ as follows:

$prop \equiv_{def} \forall X (initial(X) \rightarrow sat(X, \varphi))$

This definition can be encoded by the following two clauses:

$\gamma_1 : \ prop \leftarrow \neg negprop \qquad \gamma_2 : \ negprop \leftarrow initial(X), sat(X, not(\varphi))$

Let $P_{\mathcal{K}}$ denote the constraint logic program consisting of clauses 1–10 together with the clauses defining the predicates $initial$, $t$, $ts$, and $elem$. The program $P_{\mathcal{K}} \cup \{\gamma_1, \gamma_2\}$ is locally stratified and, hence, it has a unique perfect model denoted $M(P_{\mathcal{K}} \cup \{\gamma_1, \gamma_2\})$. The correctness of the encoding program $P_{\mathcal{K}} \cup \{\gamma_1, \gamma_2\}$ is stated by the following Theorem 1 (its proof can be found in (Fioravanti et al. 2007)).

*Theorem 1 (Correctness of Encoding)*
Let $\mathcal{K}$ be a Kripke structure, let $I$ be the set of initial states of $\mathcal{K}$, and let $\varphi$ be a CTL formula. Then, for all states $s \in I$, $\mathcal{K}, s \models \varphi$ iff $prop \in M(P_{\mathcal{K}} \cup \{\gamma_1, \gamma_2\})$.

*Example 1*
Let us consider the reactive system depicted in Figure 1, where a state $\langle X_1, X_2 \rangle$, which is a pair of rationals, is denoted by the term $s(X_1, X_2)$.

The Kripke structure $\mathcal{K}$ which models that system, is defined as follows. The initial states are given by the clause:

11. $initial(s(X_1, X_2)) \leftarrow X_1 \leq 0, \ X_2 = 0$

The transition relation $R$ is given by the clauses:

12. $t(s(X_1, X_2), s(Y_1, Y_2)) \leftarrow X_1 \geq 1, \ Y_1 = X_1, \ Y_2 = X_2 - 1$
13. $t(s(X_1, X_2), s(Y_1, Y_2)) \leftarrow X_1 \leq 2, \ Y_1 = X_1, \ Y_2 = X_2 + 1$

The predicate $ts$ is given by the clauses:

14. $ts(s(X_1, X_2), [s(Y_1, Y_2)]) \leftarrow X_1 < 1, \ Y_1 = X_1, \ Y_2 = X_2 + 1$
15. $ts(s(X_1, X_2), [s(Y_{11}, Y_{21}), s(Y_{12}, Y_{22})]) \leftarrow X_1 \geq 1, \ X_1 \leq 2,$
    $\hspace{4em} Y_{11} = X_1, \ Y_{21} = X_2 - 1, \ Y_{12} = X_1, \ Y_{22} = X_2 + 1$
16. $ts(s(X_1, X_2), [s(Y_1, Y_2)]) \leftarrow X_1 > 2, \ Y_1 = X_1, \ Y_2 = X_2 - 1$

The elementary property *negative* is given by the clause:

17. $elem(s(X_1, X_2), negative) \leftarrow X_2 < 0$

Suppose that we want to verify the property that in every initial state $s(X_1, X_2)$, where $X_1 \leq 0$ and $X_2 = 0$, the CTL formula $not(eu(true, negative))$ holds, that is, from any initial state it is impossible to reach a state $s(X_1', X_2')$ where $X_2' < 0$. By using the fact that every CTL formula of the form $not(not(\varphi))$ is equivalent to $\varphi$, this property is encoded by the following two clauses:

$\gamma_1: prop \leftarrow \neg negprop \hspace{3em} \gamma_2: negprop \leftarrow initial(X), \ sat(X, eu(true, negative))$

Note that, in this example, for the verification of *prop* the clauses defining the predicate $sat(X, af(F))$ (that is, clauses 7 and 8 of program $P_\mathcal{K}$) are not needed. Thus, clauses 14, 15, and 16, which define the predicate $ts$, are not needed either. □

Our encoding of the Kripke structure can easily be extended to provide witnesses of formulas of the form $eu(\varphi_1, \varphi_2)$ and counterexamples of formulas of the form $af(\varphi)$, as usual for model checkers of finite state systems (Clarke et al. 1999). Indeed, in order to do so, it is sufficient to add to the predicate *sat* an extra argument that recalls the sequence of states (or transitions) constructed during the verification of a given formula. For details, the reader may refer to (Fioravanti et al. 2007).

## 3 Verifying Infinite State Systems by Specializing CLP Programs

In this section we present a method for checking whether or not $prop \in M(P_\mathcal{K} \cup \{\gamma_1, \gamma_2\})$, where $P_\mathcal{K} \cup \{\gamma_1, \gamma_2\}$ is a CLP encoding of an infinite state system and *prop* is a predicate encoding the satisfiability of a given CTL formula.

As already mentioned, the proof procedures normally used in constraint logic programming, such as the extensions to CLP of SLDNF resolution and tabled resolution, very often diverge when trying to check whether or not $prop \in M(P_\mathcal{K} \cup \{\gamma_1, \gamma_2\})$ by evaluating the query *prop*. This is due to the limited ability of these proof procedures to cope with infinite failure.

Also the bottom-up construction of the perfect model $M(P_\mathcal{K} \cup \{\gamma_1, \gamma_2\})$ often diverges, because it does not take into account the information about the query *prop* to be evaluated, the initial states of the system, and the formula to be verified. Indeed, by a naive bottom-up evaluation, the clauses of $P_\mathcal{K}$ may generate infinitely many atoms of the form $sat(s, \psi)$. For instance, given a state $s_0$, an elementary property $f$ that holds in $s_0$, and an infinite sequence $\{s_i \mid i \in \mathbb{N}\}$ of distinct states such that, for every $i \in \mathbb{N}$, $t(s_{i+1}, s_i)$ holds, clauses 5 and 6 generate by bottom-up evaluation the infinitely many atoms $sat(s_i, eu(true, f))$, for every $i \in \mathbb{N}$, and the infinitely many atoms: (i) $sat(s_0, f)$, (ii) $sat(s_0, eu(true, f))$, (iii) $sat(s_0, eu(true, eu(true, f)))$, ...

We will show that the termination of the bottom-up construction of the perfect model can be improved by a prior application of program specialization. In particular, we will present our verification algorithm which consists of two phases: Phase (1), in which we specialize the program $P_\mathcal{K} \cup \{\gamma_1, \gamma_2\}$ with respect to the query *prop*, thereby deriving a new program $P_S$ whose perfect model $M(P_S)$, also denoted $M_S$, satisfies the following equivalence: $prop \in M(P_\mathcal{K} \cup \{\gamma_1, \gamma_2\})$ iff $prop \in M_S$, and Phase (2), in which we construct $M_S$ by a bottom-up evaluation. The specialization phase modifies the program $P_\mathcal{K} \cup \{\gamma_1, \gamma_2\}$ by incorporating into the specialized program $P_S$ the information about the initial states and the formula to be verified. The bottom-up evaluation of $P_S$ may terminate more often than the bottom-up evaluation of $P_\mathcal{K} \cup \{\gamma_1, \gamma_2\}$ because: (i) it avoids the generation of an infinite set of states that are unreachable from the initial states, and (ii) it generates only specialized atoms corresponding to subformulas of the formula to be verified.

**The Verification Algorithm**
*Input*: The program $P_\mathcal{K} \cup \{\gamma_1, \gamma_2\}$. *Output*: The perfect model $M_S$ of a CLP program $P_S$ such that $prop \in M(P_\mathcal{K} \cup \{\gamma_1, \gamma_2\})$ iff $prop \in M_S$.
(Phase 1)  *Specialize*$(P_\mathcal{K} \cup \{\gamma_1, \gamma_2\}, P_S)$;
(Phase 2)  *BottomUp*$(P_S, M_S)$

The *Specialize* procedure of Phase (1) consists in the iterated application of two subsidiary procedures: (i) the *Unfold* procedure, which applies the *unfolding* rule and the *clause removal* rule, and (ii) the *Generalize&Fold* procedure, which applies the *definition introduction* rule and the *folding* rule. These program transformation rules are variants, tailored to program specialization, of the usual rules for logic programs and constraint logic programs (see, for instance, (Etalle and Gabbrielli 1996; Seki 1991).

**Procedure** *Specialize*
*Input*: The program $P_\mathcal{K} \cup \{\gamma_1, \gamma_2\}$. *Output*: A stratified program $P_S$ such that $prop \in M(P_\mathcal{K} \cup \{\gamma_1, \gamma_2\})$ iff $prop \in M(P_S)$.

$P_S := \{\gamma_1\}$;   $InDefs := \{\gamma_2\}$;   $Defs := \emptyset$;
*while* there exists a clause $\gamma$ in $InDefs$
*do*  *Unfold*$(\gamma, \Gamma)$;
     *Generalize&Fold*$(Defs, \Gamma, NewDefs, \Phi)$;
     $P_S := P_S \cup \Phi$;  $InDefs := (InDefs - \{\gamma\}) \cup NewDefs$;  $Defs := Defs \cup NewDefs$;
*end-while*

The *Unfold* procedure takes as input a clause $\gamma \in InDefs$ and returns as output a set $\Gamma$ of clauses derived from $\gamma$ by one or more applications of the unfolding rule. A single application of this rule is encoded by the *UnfoldOnce* function defined below. We use the following notation. Given two atoms $A$ and $B$, we denote by $A = B$ the constraint: (i) $t_1 = u_1, \ldots, t_n = u_n$, if $A$ is of the form $p(t_1, \ldots, t_n)$ and $B$ is of the form $p(u_1, \ldots, u_n)$, for some $n$-ary predicate symbol $p$, and (ii) *false*, otherwise.

---

**Function** *UnfoldOnce*$(\gamma, A)$
Let $\gamma$ be a clause of the form $H \leftarrow c, Q, A, R$, where $A$ is an atom whose predicate is defined in $P_\mathcal{K}$. Let $\{(K_i \leftarrow c_i, B_i) \mid i = 1, \ldots, m\}$, with $m \geq 0$, be the set of (renamed apart) clauses in program $P_\mathcal{K}$ such that, for $i = 1, \ldots, m$, the constraint $(c,\ A = K_i,\ c_i)$ is satisfiable.

$\quad$ *UnfoldOnce*$(\gamma, A) = \{(K \leftarrow c, A = K_i, c_i, Q, B_i, R) \mid i = 1, \ldots, m\}$

---

At the first application of the *Unfold* procedure, the input clause $\gamma$ is the clause $\gamma_2 :\ \textit{negprop} \leftarrow \textit{initial}(X), \textit{sat}(X, \textit{not}(\varphi))$, where $\textit{initial}(X)$ and $\varphi$ encode the initial states and the formula to be verified, respectively. The *Unfold* procedure propagates the information about the initial states and the property to be verified through the Kripke structure encoded by $P_\mathcal{K}$.

---

**Procedure** *Unfold*$(\gamma, \Gamma)$
*Input*: A clause $\gamma$ in *InDefs*. *Output*: A set $\Gamma$ of clauses.

UNFOLD:
$\Gamma := \textit{UnfoldOnce}(\gamma, A)$, where $A$ is any atom in the body of $\gamma$;
*while* there exist a clause $\delta$ in $\Gamma$ and an atom $A$ in the body of $\delta$, such that $A$ is of one
$\quad$ of the following forms: (i) $\textit{initial}(s)$, (ii) $t(s_1, s_2)$, (iii) $ts(s, ss)$, (iv) $\textit{elem}(s, e)$,
$\quad$ (v) $\textit{sat}(s, e)$, where $e$ is an elementary property, (vi) $\textit{sat}(s, \textit{not}(\psi_1))$,
$\quad$ (vii) $\textit{sat}(s, \textit{and}(\psi_1, \psi_2))$, (viii) $\textit{sat}(s, \textit{ex}(\psi_1))$, (ix) $\textit{sat\_all}(ss, \psi_1)$, where $ss$ is
$\quad$ a non-variable list $\quad do\quad \Gamma := (\Gamma - \{\delta\}) \cup \textit{UnfoldOnce}(\delta, A)$
*end-while*;

REMOVE SUBSUMED CLAUSES:
*while* in $\Gamma$ there exist two distinct clauses $\delta: H \leftarrow c$ and $\eta: H \leftarrow d, G$ such that
$\quad d \sqsubseteq c$ (that is, $\eta$ is subsumed by $\delta$) $\quad do\quad \Gamma := \Gamma - \{\eta\}$
*end-while*

---

Due to the structure of the clauses in $P_\mathcal{K}$, the *Unfold* procedure terminates for every $\gamma \in \textit{InDefs}$. In particular, in order to enforce termination, every atom of the form $\textit{sat}(s, \textit{eu}(\psi_1, \psi_2))$ or $\textit{sat}(s, \textit{af}(\psi_1))$ is selected at most once during each application of the procedure.

The *Generalize&Fold* procedure takes as input the set $\Gamma$ of clauses produced by the *Unfold* procedure and introduces a set *NewDefs* of *definitions*, that is, clauses of the form $\delta: \textit{newp}(X) \leftarrow d(X), \textit{sat}(X, \psi)$, where *newp* is a new predicate. Any such clause $\delta$ represents a set of states $X$ satisfying the constraint $d(X)$ and the CTL property $\psi$, and incorporates the information which has been propagated by the *Unfold* procedure, concerning the initial state and the property to be verified. All definitions introduced by the *Generalize&Fold* procedure are stored in a set *Defs* and can be used for folding during the current or the future applications of the procedure itself. By folding the clauses in $\Gamma$ using the definitions in *Defs*∪*NewDefs*, the procedure derives a new set $\Phi$ of clauses which are added to the specialized program $P_S$. In the clauses of $P_S$, there is no reference to the predicates used in $P_\mathcal{K} \cup \{\gamma_1, \gamma_2\}$, except for *prop* and *negprop*, that is, $P_S$ provides a definition of *prop* and *negprop* in terms of the new predicates introduced by the applications of the *Generalize&Fold* procedure.

Unfortunately, an uncontrolled application of the *Generalize&Fold* procedure may lead to the introduction of infinitely many new definitions, thereby causing the nontermination of the *Specialize* procedure. In order to guarantee termination, the *Generalize&Fold* procedure may introduce new definitions which are *more general than* definitions introduced by previous applications of the procedure, where the *more general than* relation between definitions is as follows: a definition $newq(X) \leftarrow g(X), sat(X, \psi)$ is more general than the definition $newp(X) \leftarrow b(X), sat(X, \psi)$ if $b(X) \sqsubseteq g(X)$. Thus, more general definitions correspond to larger sets of states.

In order to introduce generalized definitions in a suitable way, we will extend to constraint logic programs some techniques which have been proposed for controlling generalization in *positive supercompilation* (Sørensen and Glück 1995) and *partial deduction* (Leuschel et al. 1998). The details of the *Generalize&Fold* procedure and the results stating the correctness and the termination of the *Specialize* procedure will be given in the next section.

In order to compute the perfect model $M_S$ of $P_S$ it is convenient to represent sets of ground atoms by sets of *facts*, that is, sets of (possibly non-ground) clauses of the form $H \leftarrow c$, where $H$ is an atom and $c$ is a constraint. A fact $H \leftarrow c$ represents the set of all the ground instances of $H$ that satisfy $c$. The *BottomUp* procedure constructs $M_S$ by using the *non-ground immediate consequence operator* $S_{P_S}$, instead of the usual immediate consequence operator $T_{P_S}$ (Jaffar and Maher 1994). Since program $P_S$ is *stratified* (see in Theorem 2 below), the *BottomUp* procedure processes the strata of $P_S$ from the lowest one to the highest one (that is, the stratum where the predicate *prop* occurs). For each stratum the *BottomUp* procedure computes the least fixpoint of the restriction of $S_{P_S}$ to that stratum. Since this fixpoint may be represented by an infinite set of facts, the *BottomUp* procedure may not terminate, although there is only a finite number of strata in $P_S$. In Section 5 we will see that the *BottomUp* procedure, applied after the *Specialize* procedure, terminates in many significant cases.

*Example 2*
Let us consider the program $P_\mathcal{K} \cup \{\gamma_1, \gamma_2\}$ and the query *prop* of Example 1. We have that: (i) by using a traditional Prolog system, the evaluation of *prop* does not terminate in $P_\mathcal{K} \cup \{\gamma_1, \gamma_2\}$ because *negprop* has an infinitely failed SLD tree, (ii) by using the XSB tabled logic programming system, *prop* does not terminate because infinitely many *sat* atoms are tabled, and (iii) the bottom-up construction of $M(P_\mathcal{K} \cup \{\gamma_1, \gamma_2\})$ does not terminate because of the presence of clauses 5 and 6 as we have indicated at the beginning of this section.

By applying the *Specialize* procedure to the program $P_\mathcal{K} \cup \{\gamma_1, \gamma_2\}$ (with a suitable generalization strategy, as illustrated in the next section), we derive the following specialized program $P_S$:

$\gamma_1.\ \ prop \leftarrow \neg negprop$

$\gamma'_2.\ \ negprop \leftarrow X_1 \leq 0,\ X_2 = 0,\ new1(X_1, X_2)$

$\gamma_3.\ \ new1(X_1, X_2) \leftarrow X_1 \leq 0,\ X_2 = 0,\ Y_1 = X_1,\ Y_2 = 1,\ new2(Y_1, Y_2)$

$\gamma_4.\ \ new2(X_1, X_2) \leftarrow X_1 \leq 0,\ X_2 \geq 0,\ Y_1 = X_1,\ Y_2 = X_2 + 1,\ new2(Y_1, Y_2)$

The *Specialize* procedure has propagated through the program $P_S$ the constraint $X_1 \leq 0,\ X_2 = 0$ characterizing the initial states (see clause 11 of Example 1). This constraint, in fact, occurs in clause $\gamma_3$ and its generalization $X_1 \leq 0, X_2 \geq 0$ occurs in clause $\gamma_4$. The *BottomUp* procedure computes the perfect model of $P_S$, and we get $M_S = \{prop\}$ in a finite number of steps (indeed, starting from the lowest stratum, we have that, for all $X_1, X_2$, $new2(X_1, X_2)$, $new1(X_1, X_2)$, and *negprop* are all false). Thus, the property $not(eu(true, negative))$ holds in every initial state of $\mathcal{K}$. □

## 4 Generalization Strategies

The design of a powerful generalization strategy should meet two conflicting requirements. Such a strategy, in fact, should introduce new definitions which are (i) as general as possible, so as to enforce the termination of the *Specialize* procedure, and (ii) as specific as possible, so as to retain the maximum information about the initial state and the property to be verified, and produce a program $P_S$ for which the *BottomUp* procedure terminates. In this section we present several generalization strategies for coping with those conflicting requirements. These strategies combine various techniques used in the fields of program transformation and static analysis, such as *well-binary relations*, *well-quasi orderings*, *widening*, and *convex hull* operators, and variants thereof (Cousot and Halbwachs 1978; Leuschel 2002; Leuschel et al. 1998; Peralta and Gallagher 2003; Sørensen and Glück 1995). All these strategies guarantee the termination of the *Specialize* procedure. However, since in general the verification problem is undecidable, the power and effectiveness of the different generalization strategies can only be assessed by performing experiments. The results of those experiments will be presented in the next section.

### 4.1 The Generalize&Fold Procedure

The *Generalize&Fold* procedure makes use of a tree of definitions, called *Definition Tree*, whose nodes are labelled by the clauses in $\mathit{Defs} \cup \{\gamma_2\}$. By construction there is a bijection between the set of nodes of the Definition Tree and $\mathit{Defs} \cup \{\gamma_2\}$ and, thus, we will identify each node with its label. The root of the Definition Tree is labelled by clause $\gamma_2$ (recall that $\{\gamma_2\}$ is the initial value of *InDefs*) and the children of a clause $\gamma$ in $\mathit{Defs} \cup \{\gamma_2\}$ are the clauses *NewDefs* derived after applying the procedures $\mathit{Unfold}(\gamma, \Gamma)$ and $\mathit{Generalize\&Fold}(\mathit{Defs}, \Gamma, \mathit{NewDefs}, \Phi)$. Our *Generalize&Fold* procedure is based on the combined use of a *firing relation* and a *generalization operator*. The firing relation determines *when* to generalize, while the generalization operator determines *how* to generalize.

*Definition 1 (Well-Binary Relation $\triangleleft$ and Well-Quasi Ordering $\precsim$)*
A *well-binary relation* on a set $S$ is a binary relation $\triangleleft$ such that, for every infinite sequence $e_0 e_1 \ldots$ of elements of $S$, there exist $i$ and $j$ such that $i < j$ and $e_i \triangleleft e_j$.

A *well-quasi ordering* (or *wqo*, for short) on a set $S$ is a reflexive, transitive, well-binary relation $\precsim$ on $S$. Given $e_1$ and $e_2$ in $S$, we write $e_1 \approx e_2$ if $e_1 \precsim e_2$ and $e_2 \precsim e_1$. A wqo $\precsim$ is *thin* iff for all $e \in S$, the set $\{e' \in S \mid e \approx e'\}$ is finite.

*Definition 2 (Firing Relation)*
A *firing relation* is a well-binary relation on the set $\mathcal{C}$ of constraints.

The firing relation guarantees that generalization is eventually applied and, thus, its role is similar to the one of the *whistle* algorithm (Sørensen and Glück 1995).

*Definition 3 (Generalization Operator $\ominus$)*
Let $\precsim$ be a thin wqo on the set $\mathcal{C}$ of constraints. A binary operator $\ominus$ on $\mathcal{C}$ is a *generalization operator* with respect to $\precsim$ if, for all constraints $c$ and $d$ in $\mathcal{C}$, we have: (i) $d \sqsubseteq c \ominus d$, and (ii) $c \ominus d \precsim c$. (Note that, in general, $\ominus$ is not commutative.)

The use of a thin wqo in Definition 3 guarantees that during the *Specialize* procedure each definition can be generalized a finite number of times only and, thus, the termination of the procedure is guaranteed. Definition 3 generalizes several operators proposed in the literature, such as the *most specific generalization* operator (Leuschel et al. 1998; Sørensen and Glück 1995) and the *widening* operator (Cousot and Halbwachs 1978).

---

**Procedure** *Generalize&Fold*
*Input*: (i) a set *Defs* of definitions, (ii) a set $\Gamma$ of clauses obtained from a clause $\gamma$ by the *Unfold* procedure, (iii) a firing relation $\lhd$, and (iv) a generalization operator $\ominus$.
*Output*: (i) A set *NewDefs* of new definitions, and (ii) a set $\Phi$ of folded clauses.
$NewDefs := \emptyset; \quad \Phi := \Gamma;$
*while* in $\Phi$ there exists a clause $\eta$: $H \leftarrow e, G_1, L, G_2$, where $L$ is either $sat(X, \psi)$ or
 $\neg sat(X, \psi)$ *do*

GENERALIZE:
Let $e_p(X)$ be $project(e, X)$.
1. *if* in *Defs* there exists a clause $\delta$: $newp(X) \leftarrow d(X), sat(X, \psi)$ such that
 $e_p(X) \sqsubseteq d(X)$ (modulo variable renaming)
 *then* $NewDefs := NewDefs$;
2. *elseif* there exists a clause $\alpha$ in *Defs* such that:
 (i) $\alpha$ is of the form $newq(X) \leftarrow b(X), sat(X, \psi)$, and (ii) $\alpha$ is the most
 recent ancestor of $\gamma$ in the Definition Tree such that $b(X) \lhd e_p(X)$
 *then* $NewDefs := NewDefs \cup \{newp(X) \leftarrow b(X) \ominus e_p(X), sat(X, \psi)\}$;
3. *else* $NewDefs := NewDefs \cup \{newp(X) \leftarrow e_p(X), sat(X, \psi)\}$;

FOLD:
$\Phi := (\Phi - \{\eta\}) \cup \{H \leftarrow e, G_1, M, G_2\}$, where $M$ is $newp(X)$, if $L$ is $sat(X, \psi)$,
 and $M$ is $\neg newp(X)$, if $L$ is $\neg sat(X, \psi)$

*end-while*

---

The following theorem establishes that the *Specialize* procedure always terminates and preserves the perfect model semantics. The proof of this theorem is a simple variant of the proof of Theorem 3 in (Fioravanti et al. 2007).

*Theorem 2* (*Termination and Correctness of the Specialize Procedure*)
(i) For every input program $P_\mathcal{K} \cup \{\gamma_1, \gamma_2\}$, for every firing relation $\triangleleft$, and for every generalization operator $\ominus$, the *Specialize* procedure terminates. (ii) Let $P_\mathcal{S}$ be the output program of the *Specialize* procedure. Then (ii.1) $P_\mathcal{S}$ is stratified (and thus, locally stratified), and (ii.2) $prop \in M(P_\mathcal{K})$ iff $prop \in M(P_\mathcal{S})$.

### 4.2 Firing Relations and Generalization Operators on Linear Constraints

In our verification experiments we will consider the set $Lin_k$ of constraints defined as follows. Every constraint $c \in Lin_k$ is the conjunction of $m\,(\geq 0)$ *distinct* atomic constraints $a_1, \ldots, a_m$ (and we will denote this fact by writing $c \equiv a_1, \ldots, a_m$) where, for $i = 1, \ldots, m$, (1) $a_i$ is either of the form $p_i \leq 0$ or of the form $p_i < 0$, and (2) $p_i$ is a polynomial of the form $q_0 + q_1 X_1 + \ldots + q_k X_k$, where $X_1, \ldots, X_k$ are distinct variables and $q_0, q_1, \ldots, q_k$ are integer coefficients. An equation $r = s$ stands for the conjunction of the two inequations $r \leq s$ and $s \leq r$. The constraints in $Lin_k$ are interpreted over the rationals in the usual way.

Now we present four firing relations on the set $Lin_k$. These firing relations are called *Always*, *Maxcoeff*, *Sumcoeff*, and *Homeocoeff*. They are all wqo's.

(F1) The wqo *Always*, denoted by $\precsim_A$, is the relation $Lin_k \times Lin_k$.

(F2) The wqo *Maxcoeff*, denoted by $\precsim_M$, compares the maximum absolute values of the coefficients occurring in polynomials. It is defined as follows. For any atomic constraint $a$ of the form $p < 0$ or $p \leq 0$, where $p$ is $q_0 + q_1 X_1 + \ldots + q_k X_k$, we define $maxcoeff(a)$ to be $\max\{|q_0|, |q_1|, \ldots, |q_k|\}$. Given two atomic constraints $a_1$ of the form $p_1 < 0$ and $a_2$ of the form $p_2 < 0$, we have that $a_1 \precsim_M a_2$ iff $maxcoeff(a_1) \leq maxcoeff(a_2)$. Similarly, if we are given the atomic constraints $a_1$ of the form $p_1 \leq 0$ and $a_2$ of the form $p_2 \leq 0$. Given two constraints $c_1 \equiv a_1, \ldots, a_m$, and $c_2 \equiv b_1, \ldots, b_n$, we have that $c_1 \precsim_M c_2$ iff, for $i = 1, \ldots, m$, there exists $j \in \{1, \ldots, n\}$ such that $a_i \precsim_M b_j$.

(F3) The wqo *Sumcoeff*, denoted by $\precsim_S$, compares the sum of the absolute values of the coefficients occurring in the polynomials. It is defined as follows. For any atomic constraint $a$ of the form $p < 0$ or $p \leq 0$, where $p$ is $q_0 + q_1 X_1 + \ldots + q_k X_k$, we define $sumcoeff(a)$ to be $\sum_{j=0}^{k} |q_j|$. Given two atomic constraints $a_1$ of the form $p_1 < 0$ and $a_2$ of the form $p_2 < 0$, we have that $a_1 \precsim_S a_2$ iff $sumcoeff(a_1) \leq sumcoeff(a_2)$. Similarly, if we are given the atomic constraints $a_1$ of the form $p_1 \leq 0$ and $a_2$ of the form $p_2 \leq 0$. Given two constraints $c_1 \equiv a_1, \ldots, a_m$, and $c_2 \equiv b_1, \ldots, b_n$, we have that $c_1 \precsim_S c_2$ iff, for $i = 1, \ldots, m$, there exists $j \in \{1, \ldots, n\}$ such that $a_i \precsim_S b_j$.

(F4) The wqo *Homeocoeff*, denoted by $\precsim_H$, compares sequences of absolute values of coefficients occurring in polynomials. It is an adaptation to $Lin_k$ of the *homeomorphic embedding* operator (Leuschel 2002; Leuschel et al. 1998; Sørensen and Glück 1995). The wqo $\precsim_H$ takes into account the commutativity and the associativity of addition and conjunction and it is defined as follows. Given two polynomials $p_1$ of the form $q_0 + q_1 X_1 + \ldots + q_k X_k$, and $p_2$ of the form $r_0 + r_1 X_1 + \ldots + r_k X_k$, we have that $p_1 \precsim_H p_2$ iff there exists a permutation $\langle \ell_0, \ldots, \ell_k \rangle$ of the indexes $\langle 0, \ldots, k \rangle$ such that, for $i = 0, \ldots, k$, $|q_i| \leq |r_{\ell_i}|$. Given two atomic constraints $a_1$ of the form $p_1 < 0$ and $a_2$ of the form $p_2<0$, we have that $a_1 \precsim_H a_2$ iff $p_1 \precsim_H p_2$. Similarly, if we are given the atomic constraints $a_1$ of the form $p_1\leq 0$ and $a_2$ of the form $p_2\leq 0$. Given two constraints $c_1 \equiv a_1,\ldots,a_m$, and $c_2 \equiv b_1,\ldots,b_n$ we have that $c_1 \precsim_H c_2$ iff there exist $m$ *distinct* indexes $\ell_1,\ldots,\ell_m$, with $m\leq n$, such that $a_i \precsim_H b_{\ell_i}$, for $i=1,\ldots,m$.

| $a_1$ | $a_2$ | $a_1 \precsim_A a_2$ | $a_1 \precsim_M a_2$ | $a_1 \precsim_S a_2$ | $a_1 \precsim_H a_2$ |
|---|---|---|---|---|---|
| $1-2X_1<0$ | $3+X_1<0$ | yes | yes | yes | yes |
| $2-2X_1+X_2<0$ | $1+3X_1<0$ | yes | yes | no | no |
| $1+3X_1<0$ | $2-2X_1+X_2<0$ | yes | no | yes | no |

Table 1. Examples of firing relations $\precsim_A$, $\precsim_M$, $\precsim_S$, and $\precsim_H$.

Table 1 provides some examples of the firing relations and, in particular, it shows that the relations *Maxcoeff* and *Sumcoeff* are not comparable. Figure 2(A) illustrates the containment relationships between the firing relations *Always*, *Maxcoeff*, *Sumcoeff*, and *Homeocoeff*. (The numbers appearing under each firing relation and Figure 2(B) will be explained later.) Note that a generalization operator is applied less often if it is associated with a smaller firing relation.

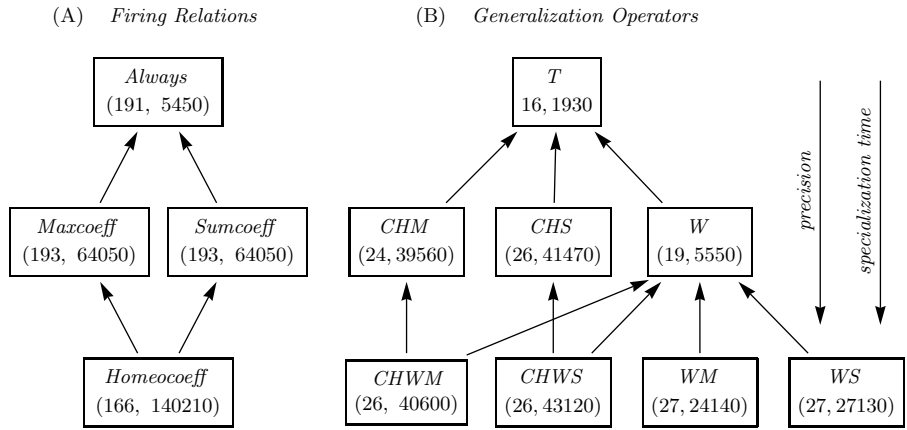

Figure 2. *Comparison of firing relations and generalization operators.*
(A) An arrow $p\to q$ from firing relation $p$ to firing relation $q$ means $p\subseteq q$. For each firing relation we have written the pair $(m,n)$, where: (i) $m$ is the number of properties verified by using that firing relation in conjunction with all generalization operators, and (ii) $n$ is the sum of the specialization times taken by using that firing relation in conjunction with all generalization operators (see Section 5).
(B) An arrow $g \to h$ from generalization operator $g$ to generalization operator $h$ means $\ominus_g \sqsubseteq \ominus_h$. For each generalization operator we have written the pair $(m,n)$, where: (i) $m$ is the total number of properties verified (see Table 3), and (ii) $n$ is the sum of the specialization times (see Table 4).

Now we present some generalization operators on $Lin_k$ which we will use in the verification examples of the next section. For defining these operators we will use the relations $\precsim_M$, $\precsim_S$, and $\precsim_H$, which are thin wqo's on $Lin_k$. On the contrary, the wqo $\precsim_A$ is *not* thin and it cannot be used for defining generalization operators.

(G1) Given any two constraints $c$ and $d$, the operator *Top*, denoted $\ominus_T$, returns *true*. It can be shown that *Top* is a generalization operator with respect to any of the thin wqo's $\precsim_M$, $\precsim_S$, and $\precsim_H$. Since the *Top* operator forgets all information about its operands, it often performs an over-generalization and produces poorly specialized programs (see the experimental evaluation in Section 5).

(G2) Given any two constraints $c \equiv a_1, \ldots, a_m$, and $d$, the operator *Widen*, denoted $\ominus_W$, returns the constraint $a_{i1}, \ldots, a_{ir}$, such that $\{a_{i1}, \ldots, a_{ir}\} = \{a_h \mid 1 \leq h \leq m$ and $d \sqsubseteq a_h\}$. Thus, *Widen* returns all atomic constraints of $c$ that are entailed by $d$ (see (Cousot and Halbwachs 1978) for a similar widening operator used in static program analysis). The operator $\ominus_W$ is a generalization operator w.r.t. any of the thin wqo's $\precsim_M$, $\precsim_S$, and $\precsim_H$.

(G3) Given any two constraints $c \equiv a_1, \ldots, a_m$, and $d \equiv b_1, \ldots, b_n$, the operator *WidenMax*, denoted $\ominus_{WM}$, returns the conjunction $a_{i1}, \ldots, a_{ir}, b_{j1}, \ldots, b_{js}$, where: (i) $\{a_{i1}, \ldots, a_{ir}\} = \{a_h \mid 1 \leq h \leq m$ and $d \sqsubseteq a_h\}$, and (ii) $\{b_{j1}, \ldots, b_{js}\} = \{b_k \mid 1 \leq k \leq n$ and $b_k \precsim_M c\}$. The operator *WidenSum*, denoted $\ominus_{WS}$, is defined like *WidenMax*, with $\precsim_M$ replaced by $\precsim_S$. The operators $\ominus_{WM}$ and $\ominus_{WS}$ are generalization operators w.r.t. the thin wqo's $\precsim_M$ and $\precsim_S$, respectively.

The operators *WidenMax* and *WidenSum* are similar to *Widen* but, together with the atomic constraints of $c$ that are entailed by $d$, they also return the conjunction of a subset of the atomic constraints of $d$. Note that the operator *WidenHomeo*, denoted $\ominus_{WH}$, which is defined like *WidenMax*, with $\precsim_M$ replaced by $\precsim_H$, is *not* a generalization operator w.r.t. $\precsim_H$. Indeed, the constraint $c \ominus_{WH} d$ may contain more atomic constraints than $c$ and, thus, it may not be the case that $(c \ominus_{WH} d) \precsim_H c$.

Next we define some generalization operators by using the *convex hull* operator, which sometimes is used in the static program analysis (Cousot and Halbwachs 1978). The *convex hull* of two constraints $c$ and $d$ in $Lin_k$, denoted by $ch(c, d)$, is the least (w.r.t. the $\sqsubseteq$ ordering) constraint $h$ in $Lin_k$ such that $c \sqsubseteq h$ and $d \sqsubseteq h$. (Note that $ch(c, d)$ is unique up to equivalence of constraints.)

(G4) Given any two constraints $c$ and $d$, let $ch(c, d)$ be of the form $b_1, \ldots, b_n$. The operator *CHMax*, denoted $\ominus_{CHM}$, returns the conjunction $b_{j1}, \ldots, b_{js}$, such that $\{b_{j1}, \ldots, b_{js}\} = \{b_k \mid 1 \leq k \leq n$ and $b_k \precsim_M c\}$. The operator *CHSum*, denoted $\ominus_{CHS}$, is defined like *CHMax*, with $\precsim_M$ replaced by $\precsim_S$. The operators $\ominus_{CHM}$ and $\ominus_{CHS}$ are generalization operators w.r.t. the thin wqo's $\precsim_M$ and $\precsim_S$, respectively.

Both *CHMax* and *CHSum* return the conjunction of a subset of the atomic constraints of $ch(c, d)$. Note that if in the definition of *CHMax* we replace $\precsim_M$ by $\precsim_H$, we get an operator which is not a generalization operator.

(G5) Given any two constraints $c$ and $d$, we define the operator *CHWidenMax*, denoted $\ominus_{CHWM}$, as follows: $c \ominus_{CHWM} d = c \ominus_{WM} ch(c, d)$. Similarly, we define the operator *CHWidenSum*, denoted $\ominus_{CHWS}$, as follows: $c \ominus_{CHWS} d = c \ominus_{WS} ch(c, d)$.

| | | | |
|---|---|---|---|
| $c$ | $-X_1 \leq 0,\ -2+X_1 \leq 0$ | $1-X_1 \leq 0,\ -2+X_1 \leq 0$ | $1-X_1 \leq 0, -1+X_1 \leq 0,$ $X_2 \leq 0,\ -X_2 \leq 0$ |
| $d$ | $2-X_1 \leq 0,\ 1-X_2 \leq 0$ | $-X_1 \leq 0$ | $X_1 \leq 0,\ -X_1 \leq 0,$ $2-X_2 \leq 0,\ -2+X_2 \leq 0$ |
| $c \ominus_W d$ | $-X_1 \leq 0$ | $true$ | $-1+X_1 \leq 0, -X_2 \leq 0$ |
| $c \ominus_{WM} d$ | $2-X_1 \leq 0,\ 1-X_2 \leq 0$ | $-X_1 \leq 0$ | $1-X_1 \leq 0,\ -1+X_1 \leq 0,$ $-X_2 \leq 0$ |
| $c \ominus_{CHM} d$ | $-X_1 \leq 0$ | $-X_1 \leq 0$ | $-X_2 \leq 0$ |
| $c \ominus_{CHWM} d$ | $-X_1 \leq 0$ | $-X_1 \leq 0$ | $-1+X_1 \leq 0, -X_2 \leq 0$ |

Table 2. Examples of application of generalization operators.

The operators $\ominus_{CHWM}$ and $\ominus_{CHWS}$ are generalization operators w.r.t. the thin wqo's $\precsim_M$ and $\precsim_S$, respectively.

Both *CHWidenMax* and *CHWidenSum* return the conjunction of a subset of the atomic constraints of $c$ and a subset of the atomic constraints of $ch(c, d)$.

Note that some other combinations of the widening and convex hull operators would not yield new generalization operators. Indeed, for all constraints $c$ and $d$, we have that: (i) $c \ominus_T ch(c, d) = c \ominus_T d$, (ii) $c \ominus_W ch(c, d) = c \ominus_W d$, (iii) $c \ominus_{CHM} ch(c, d) = c \ominus_{CHM} d$, and (iv) $c \ominus_{CHS} ch(c, d) = c \ominus_{CHS} d$.

It can be shown that the generalization operators defined at points (G1)–(G5) above are pairwise distinct. Table 2 shows some examples of application of generalization operators.

In order to compare our generalization operators we extend the $\sqsubseteq$ partial ordering on constraints to a partial ordering, also denoted $\sqsubseteq$, on generalization operators, as follows: $\ominus_1 \sqsubseteq \ominus_2$ (and we say that $\ominus_1$ is less general than $\ominus_2$) iff, for all constraints $c$ and $d$, $(c \ominus_1 d) \sqsubseteq (c \ominus_2 d)$. Figure 2(B) shows the relationships between generalization operators. (The numbers appearing under each generalization operator will be explained in Section 5.) The operators not connected by any sequence of arrows are not comparable w.r.t. $\sqsubseteq$.

## 5 Experimental Evaluation

In this section we present the results of the experiments we have performed on several examples of verification of infinite state reactive systems. We have implemented the verification algorithm presented in Section 2 using MAP, an experimental system for transforming constraint logic programs (MAP 2011). The MAP system is implemented in SICStus Prolog 3.12.8 and uses the `clpq` library to operate on constraints.

We have considered the following *mutual exclusion* protocols and we have verified some of their properties. (i) *Bakery* (Delzanno and Podelski 2001): we have verified safety (that is, mutual exclusion) and liveness (that is, starvation freedom) in the case of two processes, and safety in the case of three processes; (ii) *MutAst* (Lesens and Saïdi 1997): we have verified safety in the case of two processes; (iii) *Peterson* (Bardin et al. 2008): we have verified safety in the case of $N$ ($\geq 2$) processes by considering a *counting abstraction* of the protocol (Delzanno 2003); and (iv) *Ticket* (Delzanno and Podelski 2001): we have verified safety and liveness in the case of two processes.

We have also verified safety properties of the following *cache coherence* protocols: (v) *Berkeley RISC*, (vi) *DEC Firefly*, (vii) *IEEE Futurebus+*, (viii) *Illinois University*, (ix) *MESI*, (x) *MOESI*, (xi) *Synapse N+1*, and (xii) *Xerox PARC Dragon*. We have considered *parameterized* versions of the protocols (v)–(xii), that is, protocols designed for an arbitrary number of processors. We have applied our verification method to the counting abstractions described in (Delzanno 2003).

Then we have verified safety properties of the following systems. (xiii) *Barber* (Bultan 2000): we have considered a parameterized version of this protocol with a single barber process and an arbitrary number of customer processes; (xiv) *Bounded Buffer* and *Unbounded Buffer*: we have considered protocols for two producers and two consumers which communicate via a bounded and an unbounded buffer, respectively (the encodings of these protocols are taken from (Delzanno and Podelski 2001)); (xv) *Consprodjava*, which is (a counting abstraction of) a producer–consumer Java program realized using threads: we have verified that for any number of threads there is no deadlock (Bardin et al. 2008); (xvi) *CSM* is a central server model described in (Delzanno et al. 1999); (xvii) *Consistency*, which is a directory-based consistency protocol for client–server distributed systems (proposed by Steven German) (Bardin et al. 2008): we have considered two versions of the system and we have verified that mutual exclusion is preserved for any number of processes; (xviii) *Insertion Sort* and *Selection Sort*: we have considered the problem of checking array bounds of these two sorting algorithms, parameterized w.r.t. the size of the array, as presented in (Delzanno and Podelski 2001); (xix) *Office Light Control* (Yavuz-Kahveci and Bultan 2009) is a protocol for controlling how office lights are switched on and off, depending on room occupancy; (xx) *Reset Petri Net* is a Petri Net augmented with *reset arcs*: we have considered a reachability problem for a net which is a variant of one presented in (Leuschel and Lehmann 2000); (xxi) *Kanban* is a Petri Net modelling a concurrent production system (Bardin et al. 2008): we have verified that the value of certain control variables are bound within some specified limits; (xxii) *Train* is an encoding of a control system for speed regulation of subway trains (Bardin et al. 2008): we have verified that a train is never too early or too late with respect to its expected arrival time.

Tables 3 and 4 show the results of running the MAP system on the above examples by using the firing relation *Always* in conjunction with each of the eight generalization operators introduced in Section 4. In particular, Table 3 reports, for each example, the total verification time, that is, the time taken by the *Verification* algorithm, if it terminates, and Table 4 reports the *specialization time*, that is, the time taken by the *Specialize* procedure only. For a meaningful comparison between total specialization times, we have omitted from Table 4 the times relative to the *Consprodjava* example, for which the *Specialize* procedure does not terminate when using some of the generalization operators.

|  | $T$ | $W$ | $CHM$ | $CHS$ | $CHWM$ | $CHWS$ | $WM$ | $WS$ |
|---|---:|---:|---:|---:|---:|---:|---:|---:|
| Bakery2 (safety) | 80 | 150 | 30 | 30 | 40 | 50 | 30 | 20 |
| Bakery2 (liveness) | $\infty$ | $\infty$ | 110 | 100 | 130 | 120 | 90 | 60 |
| Bakery3 (safety) | 3940 | 4900 | 430 | 430 | 460 | 460 | 170 | 170 |
| MutAst | 2330 | 320 | 390 | 400 | 420 | 440 | 80 | 160 |
| Peterson | $\infty$ | $\infty$ | 860 | 870 | 1380 | 1410 | 190 | 220 |
| Ticket (safety) | 20 | 30 | 30 | 20 | 20 | 10 | 20 | 20 |
| Ticket (liveness) | 110 | 100 | 100 | 100 | 80 | 90 | 80 | 110 |
| Berkeley RISC | 30 | 30 | 180 | 170 | 210 | 200 | 30 | 30 |
| DEC Firefly | 70 | 130 | 200 | 130 | 310 | 320 | 20 | 30 |
| IEEE Futurebus+ | 16380 | 47570 | $\infty$ | 47120 | 75860 | 47630 | 110 | 2460 |
| Illinois University | 100 | 70 | 50 | 60 | 50 | 50 | 10 | 30 |
| MESI | 70 | 40 | 250 | 250 | 130 | 130 | 30 | 30 |
| MOESI | 100 | 150 | 330 | 170 | 120 | 170 | 40 | 60 |
| Synapse N+1 | 10 | 20 | 10 | 30 | 30 | 30 | 20 | 20 |
| Xerox PARC Dragon | 50 | 60 | 220 | 220 | 280 | 270 | 30 | 30 |
| Barber | $\infty$ | 28440 | 2000 | 2050 | 2530 | 2560 | 1160 | 1220 |
| Bounded Buffer | 30 | 360 | 9490 | 9570 | 5800 | 5840 | 3580 | 3580 |
| Unbounded Buffer | $\infty$ | $\infty$ | 410 | 400 | 420 | 420 | 3810 | 3810 |
| Consprodjava | $\infty$ | $\infty$ | $\infty$ | $\infty$ | $\infty$ | $\infty$ | 25300 | $\infty$ |
| CSM | $\infty$ | $\infty$ | 3820 | 3880 | 4830 | 4860 | 6410 | 6540 |
| Consistency v1 | $\infty$ | $\infty$ | 410 | 450 | 780 | 780 | 70 | 60 |
| Consistency v2 | $\infty$ | 70 | 110 | 130 | 220 | 260 | 40 | 60 |
| Insertion Sort | 80 | 70 | 130 | 120 | 160 | 170 | 100 | 90 |
| Selection Sort | $\infty$ | $\infty$ | $\infty$ | 160 | 230 | 180 | $\infty$ | 180 |
| Office Light Control | 50 | 40 | 50 | 50 | 50 | 50 | 50 | 50 |
| Reset Petri Net | $\infty$ | $\infty$ | $\infty$ | $\infty$ | $\infty$ | $\infty$ | 20 | 20 |
| Kanban | $\infty$ | $\infty$ | 15630 | 15800 | 17790 | 18040 | 8130 | 8000 |
| Train | $\infty$ | 1440 | 3420 | 6290 | 3680 | 6650 | 30900 | 57260 |

Table 3. Verification times for the MAP system. For each example we show the *total verification time* (Phases 1 and 2) obtained by using the firing relation *Always* in conjunction with the generalization operators: $\ominus_T$, $\ominus_W$, $\ominus_{CHM}$, $\ominus_{CHS}$, $\ominus_{CHWM}$, $\ominus_{CHWS}$, $\ominus_{WM}$, and $\ominus_{WS}$. Times are expressed in milliseconds (ms). '$\infty$' means no answer within 100 seconds.

Let us compare the various generalization operators with respect to *precision*, that is, with respect to the number of properties verified. As expected, we have that precision increases when we use less general generalization operators, that is, precision is anti-monotonic with respect to the $\sqsubseteq$ relation (precision increases when going down in Figure 2(B)). This anti-monotonicity is explained by the fact that the use of less general generalization operators may produce specialized programs that better exploit the information about both the initial state and the property to be verified.

Let us now compare the various generalization operators with respect to the specialization time. We have that specialization times increase when we use less general generalization operators, that is, specialization time is anti-monotonic with respect to the $\sqsubseteq$ relation (specialization time increases when going down in Figure 2(B)). This is due to the fact that less general generalization operators may introduce more definitions and, therefore, the specialization phase may take more time. Note also that the generalization operators that use the convex hull operators (that is, $\ominus_{CHM}$, $\ominus_{CHS}$, $\ominus_{CHWM}$, and $\ominus_{CHWS}$) exhibit higher specialization times than the ones that do not. This is due to the extra cost of computing the convex hull which, however, does not always correspond to an increase of precision.

|  | $T$ | $W$ | $CHM$ | $CHS$ | $CHWM$ | $CHWS$ | $WM$ | $WS$ |
|---|---|---|---|---|---|---|---|---|
| Bakery2 (safety) | 20 | 60 | 30 | 30 | 40 | 50 | 30 | 20 |
| Bakery2 (liveness) | 80 | 120 | 80 | 70 | 90 | 80 | 60 | 30 |
| Bakery3 (safety) | 690 | 610 | 410 | 410 | 440 | 440 | 160 | 160 |
| MutAst | 210 | 280 | 360 | 370 | 400 | 420 | 70 | 140 |
| Peterson | 20 | 250 | 850 | 870 | 1370 | 1400 | 190 | 220 |
| Ticket (safety) | 20 | 30 | 30 | 20 | 20 | 10 | 20 | 20 |
| Ticket (liveness) | 70 | 60 | 60 | 60 | 40 | 50 | 40 | 70 |
| Berkeley RISC | 20 | 20 | 150 | 140 | 180 | 170 | 30 | 30 |
| DEC Firefly | 20 | 60 | 100 | 70 | 150 | 160 | 20 | 30 |
| IEEE Futurebus+ | 30 | 230 | 1540 | 300 | 1110 | 290 | 110 | 250 |
| Illinois University | 30 | 50 | 40 | 50 | 50 | 50 | 10 | 30 |
| MESI | 20 | 30 | 150 | 150 | 120 | 120 | 30 | 30 |
| MOESI | 30 | 60 | 140 | 80 | 100 | 80 | 40 | 50 |
| Synapse N+1 | 10 | 10 | 10 | 20 | 30 | 20 | 20 | 10 |
| Xerox PARC Dragon | 20 | 30 | 190 | 190 | 260 | 250 | 30 | 30 |
| Barber | 400 | 1590 | 1870 | 1920 | 2400 | 2430 | 1130 | 1190 |
| Bounded Buffer | 10 | 140 | 9480 | 9560 | 5790 | 5830 | 2070 | 2070 |
| Unbounded Buffer | 20 | 100 | 410 | 400 | 410 | 410 | 350 | 350 |
| CSM | 30 | 450 | 3810 | 3870 | 4820 | 4840 | 6350 | 6480 |
| Consistency v1 | 20 | 90 | 410 | 450 | 770 | 770 | 70 | 60 |
| Consistency v2 | 20 | 70 | 110 | 130 | 220 | 260 | 30 | 50 |
| Insertion Sort | 20 | 50 | 130 | 120 | 150 | 160 | 100 | 90 |
| Selection Sort | 30 | 70 | 190 | 160 | 220 | 180 | 780 | 170 |
| Office Light Control | 40 | 30 | 40 | 40 | 40 | 40 | 40 | 40 |
| Reset Petri Net | 10 | 10 | 10 | 10 | 10 | 10 | 10 | 10 |
| Kanban | 20 | 1000 | 15590 | 15750 | 17740 | 18010 | 8070 | 7940 |
| Train | 20 | 50 | 3370 | 6230 | 3630 | 6590 | 4280 | 7560 |
| TOTAL | 1930 | 5550 | 39560 | 41470 | 40600 | 43120 | 24140 | 27130 |

Table 4. Specialization times for the MAP system. For each example we show the *specialization time* (Phases 1 only) obtained by using the firing relation *Always* in conjunction with the generalization operators: $\ominus_T$, $\ominus_W$, $\ominus_{CHM}$, $\ominus_{CHS}$, $\ominus_{CHWM}$, $\ominus_{CHWS}$, $\ominus_{WM}$, and $\ominus_{WS}$. Times are expressed in milliseconds (ms).

If we compare the various generalization operators by using them in conjunction with each firing relation *Maxcoeff*, *Sumcoeff*, and *Homeocoeff*, instead of *Always*, we get similar anti-monotonicity results (not shown here) for precision and specialization times.

Let us now compare the firing relations *Always*, *Maxcoeff*, *Sumcoeff*, and *Homeocoeff*. We may expect that a firing relation that determines fewer generalization steps, also determines the introduction of more definitions and, therefore, we may expect that both precision and specialization time are anti-monotonic with respect to $\subseteq$ (they increase when going down in Figure 2(A)). This anti-monotonicity is, in fact, observed in our experiments except for the case of the *Homeocoeff* firing relation (see Figure 2(A)). This is explained by the fact that the specialization times obtained by using the *Homeocoeff* firing relation are very high and, therefore, the execution of the *Specialize* procedure is often longer than the time limit of 100 seconds we have assumed as a time out. Note also that the modest increase of precision from *Always* to *Maxcoeff* or *Sumcoeff* (from 191 to 193) is paid by a considerable increase of specialization time (from 5450 ms to 64050 ms).

In summary, if we consider the balance between precision and time, the generalization strategies that use *Always* as firing relation and either $\ominus_{WM}$ or $\ominus_{WS}$ as generalization operators, outperform all the others. In particular, the generalization strategies based on the homeomorphic embedding as a firing relation (that is, *Homeocoeff*) and the convex hull operator (that is, $\ominus_{CHM}$, $\ominus_{CHS}$, $\ominus_{CHWM}$, and $\ominus_{CHWS}$) turn out not to be the best strategies in our examples.

In order to compare the implementation of our verification method using MAP with other constraint-based model checking tools for infinite state systems available in the literature, we have done the verification examples described in Table 3 on the following systems as well: (i) ALV (Yavuz-Kahveci and Bultan 2009), which combines BDD-based symbolic manipulation for boolean and enumerated types, with a solver for linear constraints on integers, (ii) DMC (Delzanno and Podelski 2001), which computes (approximated) least and greatest fixpoints of CLP(R) programs, and (iii) HyTech (Henzinger et al. 1997), a model checker for hybrid systems which handles constraints on reals. All experiments with the MAP, ALV, DMC, and HyTech systems have been performed on an Intel Core 2 Duo E7300 2.66GHz under the Linux operating system. Table 5 reports the results obtained by using various options available in those verification systems.

Table 5 indicates that, in terms of precision, MAP with either the *WM* or the *WS* generalization operator is the best system (27 properties verified out of 28), followed by ALV with the *default* option (20 out of 28), DMC with the *A* (abstraction) option (19 out of 28), and HyTech with the *Bw* (backward reachability) option (18 out of 28).

In order to compare the systems in terms of verification times, now we consider the options that give the best precision, that is, MAP with *WM*, ALV with *default*, DMC with *A*, and HyTech with *Bw*. Then we compare MAP to every other system by computing the average verification time over the set of examples where the systems terminate. We have that MAP has better average time than ALV (2343 ms and 9816 ms average time, respectively, over the 20 examples where both systems terminate), and MAP has also better average time than DMC (298 ms and 819 ms, respectively, over 19 examples). However, MAP has a slightly worse average time than HyTech (519 ms and 331 ms, respectively, over 18 examples). This is explained by the fact that HyTech with the *Bw* option tries to verify a safety property with a very simple strategy, that is, by constructing the reachability set backwards from the property to be verified, while MAP applies much more sophisticated techniques. Note also that the average verification times are affected by the peculiar behaviour on some specific examples. For instance, in the Bounded Buffer and the Barber examples the MAP system has longer verification times with respect to HyTech, because these examples can be easily verified by backward reachability and, thus, the MAP specialization phase, which propagates the information about the initial state, is redundant. On the opposite side, MAP is more efficient than HyTech in the IEEE Futurebus+ and Bakery3 examples.

## 6 Conclusions

This paper extends earlier work presented in (Fioravanti et al. 2001; Fioravanti et al. 2011). We have presented a specialization-based method for the verification of CTL properties of infinite state reactive systems. Our method consists of two phases: in Phase (1) a CLP specification of the reactive system is specialized w.r.t. the initial state and the temporal property to be verified, and in Phase (2) the perfect model of the specialized program is constructed in a bottom-up way.

For Phase (1) we have focused on the generalization strategy which is applied during program specialization and which often determines the quality of the specialized program. We have considered various generalization strategies that employ different firing relations, for deciding *when* to apply generalization, and generalization operators, for deciding *how* to generalize. The notions of firing relation and generalization operator extend to CLP the notions of *whistle* algorithm and *most specific generalization* operator, respectively, which have been proposed for positive supercompilation (Sørensen and Glück 1995) and partial deduction (Leuschel et al. 1998). For defining firing relations we have extended well-binary relations already considered in the program specialization literature, such as the homeomorphic embedding relation (Leuschel 2002; Leuschel et al. 1998; Sørensen and Glück 1995), and for defining generalization operators we have adapted notions from the area of static program analysis, such as the ones of widening and convex hull (Cousot and Halbwachs 1978). We have also introduced some new notions based on maximal coefficients and sums of coefficients of polynomials.

We have applied our verification method to several examples of infinite state systems taken from the literature, and we have compared the results in terms of precision and efficiency (that is, the number of properties which have been verified and the time taken for verification). On the basis of our experimental results we have found that some generalization strategies outperform all the others. In particular, the strategies based on maximal coefficients and sums of coefficients appear to have the best balance between precision and efficiency.

Then, we have applied other tools for the verification of infinite state systems (in particular, ALV (Yavuz-Kahveci and Bultan 2009), DMC (Delzanno and Podelski 2001), and HyTech (Henzinger et al. 1997)) to the same set of examples. The experiments show that our specialization-based verification system is quite competitive, especially in terms of precision.

Our approach is closely related to other verification methods for infinite state systems based on the specialization of (constraint) logic programs (Leuschel and Lehmann 2000; Leuschel and Massart 2000; Peralta and Gallagher 2003). However, unlike the ap-proach proposed in (Leuschel and Lehmann 2000; Leuschel and Massart 2000) we use constraints, which give us very powerful ways for dealing with infinite sets of states. The specialization-based verification method presented in (Peralta and Gallagher 2003) consists of one phase only, incorporating top-down query directed specialization and bottom-up answer propagation. That method is restricted to definite constraint logic programs and makes use of a generalization technique which combines widening and convex hull computations for ensuring termination. However, in (Peralta and Gallagher 2003) only two examples of verification have been presented (the Bakery protocol and a Petri net) and no verification times are reported and, thus, it is hard to make an experimental comparison of that method with our method.

|  | MAP | | ALV | | | | DMC | | HyTech | |
| EXAMPLE | WM | WS | default | A | F | L | noAbs | Abs | Fw | Bw |
| --- | --- | --- | --- | --- | --- | --- | --- | --- | --- | --- |
| Bakery2 (safety) | 30 | 20 | 20 | 30 | 90 | 30 | 10 | 30 | $\infty$ | 20 |
| Bakery2 (liveness) | 90 | 60 | 30 | 30 | 90 | 30 | 60 | 70 | $\times$ | $\times$ |
| Bakery3 (safety) | 170 | 170 | 580 | 570 | $\infty$ | 600 | 460 | 3090 | $\infty$ | 360 |
| MutAst | 80 | 160 | 1460 | 1000 | 220 | 1510 | 150 | 1370 | 70 | 130 |
| Peterson N | 190 | 220 | 71690 | $\bot$ | $\infty$ | $\infty$ | $\infty$ | $\infty$ | 70 | $\infty$ |
| Ticket (safety) | 20 | 20 | $\infty$ | 80 | 30 | $\infty$ | $\infty$ | 60 | $\infty$ | $\infty$ |
| Ticket (liveness) | 80 | 110 | $\infty$ | 230 | 40 | $\infty$ | $\infty$ | 220 | $\times$ | $\times$ |
| Berkeley RISC | 30 | 30 | 10 | $\bot$ | 20 | 60 | 30 | 30 | $\infty$ | 20 |
| DEC Firefly | 20 | 20 | 10 | $\bot$ | 20 | 80 | 50 | 80 | $\infty$ | 20 |
| IEEE Futurebus+ | 110 | 2460 | 320 | $\bot$ | $\infty$ | 670 | 4670 | 9890 | $\infty$ | 380 |
| Illinois University | 10 | 30 | 10 | $\bot$ | $\infty$ | 140 | 70 | 110 | $\infty$ | 20 |
| MESI | 30 | 30 | 10 | $\bot$ | 20 | 60 | 40 | 60 | $\infty$ | 20 |
| MOESI | 40 | 60 | 10 | $\bot$ | 40 | 100 | 50 | 90 | $\infty$ | 10 |
| Synapse N+1 | 20 | 20 | 10 | $\bot$ | 10 | 30 | 10 | 10 | $\infty$ | 10 |
| Xerox PARC Dragon | 30 | 30 | 20 | $\bot$ | 40 | 340 | 70 | 120 | $\infty$ | 20 |
| Barber | 1160 | 1220 | 340 | $\bot$ | 90 | 360 | 140 | 230 | $\infty$ | 90 |
| Bounded Buffer | 3580 | 3580 | 10 | 10 | $\infty$ | 20 | 20 | 30 | $\infty$ | 10 |
| Unbounded Buffer | 3810 | 3810 | 10 | 10 | 40 | 40 | $\infty$ | $\infty$ | $\infty$ | 20 |
| Consprodjava | 25300 | $\infty$ | $\infty$ | $\infty$ | $\infty$ | $\infty$ | $\infty$ | $\infty$ | $\infty$ | $\infty$ |
| CSM | 6410 | 6540 | 79490 | $\bot$ | $\infty$ | $\infty$ | $\infty$ | $\infty$ | $\infty$ | $\infty$ |
| Consistency v1 | 70 | 60 | $\infty$ | $\bot$ | $\infty$ | $\infty$ | $\infty$ | $\infty$ | $\infty$ | 2030 |
| Consistency v2 | 40 | 60 | $\infty$ | $\bot$ | 40 | $\infty$ | $\infty$ | $\infty$ | $\infty$ | 2790 |
| Insertion Sort | 100 | 90 | 40 | 60 | $\infty$ | 70 | 30 | 80 | $\infty$ | 10 |
| Selection Sort | $\infty$ | 180 | $\infty$ | 390 | $\infty$ | $\infty$ | $\infty$ | $\infty$ | $\infty$ | $\infty$ |
| Office Light Control | 50 | 50 | 20 | 20 | 30 | 20 | 10 | 10 | $\infty$ | $\infty$ |
| Reset Petri Net | 20 | 20 | $\infty$ | $\bot$ | $\infty$ | 10 | 10 | 10 | $\infty$ | 10 |
| Kanban | 8130 | 8000 | $\infty$ | $\infty$ | $\infty$ | $\infty$ | $\infty$ | $\infty$ | 700 | $\infty$ |
| Train | 30900 | 57260 | 42240 | $\bot$ | $\infty$ | 30 | $\infty$ | $\infty$ | $\infty$ | $\infty$ |
| no. of verified properties | 27 | 27 | 20 | 11 | 15 | 19 | 17 | 19 | 3 | 18 |

Table 5. Comparison of the MAP, ALV, DMC, and HyTech verification systems. Times are expressed in milliseconds. (i) '$\bot$' means termination with the answer: 'Unable to verify'. (ii) '$\infty$' means 'No answer' within 100 seconds. (iii) '$\times$' means that the test has not been performed (HyTech has no built-in for checking liveness). For the MAP system we show the total verification times with the *WM* and *WS* generalization operators (see the last two columns of Table 3). For the ALV system we show the times for four options: *default*, *A* (with approximate backward fixpoint computation), *F* (with approximate forward fixpoint computation), and *L* (with computation of loop closures for accelerating reachability). For the DMC system we show the times for two options: *noAbs* (without abstraction) and *Abs* (with abstraction). For the HyTech system we show the times for two options: *Fw* (forward reachability) and *Bw* (backward reachability).

Another approach based on program transformation for verifying parameterized systems has been presented in (Roychoudhury et al. 2000). It is an approach based on unfold/fold transformations which are more general than the ones used by us. However, the strategy for guiding the unfold/fold rules proposed in (Roychoudhury et al. 2000) works in fully automatic mode in a small set of examples only.

Finally, we would like to mention that our verification method can be regarded as complementary with respect to the methods for the verification of infinite state systems based on abstraction (Abdulla et al. 2009; Banda and Gallagher 2010; Clarke et al. 1994; Dams et al. 1997; Delzanno and Podelski 2001; Geeraerts et al. 2006; Godefroid et al. 2001). These methods work by constructing approximations of the set of reachable states that satisfy a given property. In contrast, the specialization technique applied during Phase (1) of our method, transforms the program for computing sets of states, but it does not change the set of states satisfying the property of interest. Moreover, during Phase (2) we perform an exact computation of the perfect model of the transformed program.

Further enhancements of infinite state verification could be achieved by combining program specialization and abstraction. In particular, an extension of our method could be done by replacing the bottom-up, exact computation of the perfect model performed in Phase (2), by an approximated computation in the style of (Banda and Gallagher 2010; Delzanno and Podelski 2001). However, this extension would require the computation of both over-approximations and under-approximations of models, because of the presence of negation. An interesting direction for future research is the study of how to combine in the best way, both in terms of precision and efficiency, the verification techniques based on program specialization and the ones based on abstraction.

## Acknowledgements

We thank the anonymous referees of CILC 2010 and LOPSTR 2010 for very valuable comments on earlier versions of this paper. We also thank Laurent Fribourg for stimulating discussions and John Gallagher for giving us the code of some of the systems considered in Section 5. This work has been partially supported by ERCIM, by PRIN-MIUR, and by a joint project between CNR (Italy) and CNRS (France).